\documentclass[referee]{aa}
\usepackage{graphicx}
\begin{document}
\title{
Asteroseismic Signatures of Helium gradients in Main-Sequence A Stars ; 
Application to the roAp Star HD60435
}
\author{S. Vauclair\inst{1}
 \and S. Th\'eado\inst{1,2}
}
\offprints{S. Vauclair}
\institute{ Universit\'e Paul Sabatier, Observatoire Midi-Pyr\'en\'ees,
  CNRS/UMR5572,
 14 av. E. Belin, 31400 Toulouse, France
\and  Centro de Astrofisica da Universidade do Porto, 
rua das Estrelas, 4150-762 Porto, Portugal
}
\titlerunning{ Asteroseismology of A Stars }
\authorrunning{Vauclair and Th\'eado}
\date{Received; accepted}
\abstract{
Asteroseismology is found to be a excellent tool for detecting 
diffusion-induced
helium gradients inside main-sequence A stars. Models have been computed
for 1.6 and 2.0 M$_{\odot}$ stars with pure helium diffusion, at different
ages, so that the helium gradient lies at different depths inside the star.
The adiabatic oscillation frequencies have been analysed and compared
with those of a model without diffusion. Clear signatures of the
diffusion-induced helium gradient are found in the so-called
``second differences" : these frequency differences present modulations
due to the partial reflexion of the sound waves on the layer
where the helium gradient occurs.
A tentative application to the
roAp star HD60435, which presents enough
detected oscillation frequencies for the test to be possible,
is very encouraging. The results suggest the presence of 
a helium gradient inside the star, which is consistent with  
the idea that the triggering of the oscillations
is due to the hydrogen $\kappa$-mechanism. 
}

\maketitle

\section{Introduction}

It is now widely recognized that
the abundance anomalies observed in peculiar A stars are basically due
to element 
diffusion. Those for which the radiative acceleration is larger than
gravity 
(like most of the metals) are pushed upwards while the others,such as 
helium, diffuse downwards (Michaud \cite{michaud70}, Vauclair {\&} Vauclair \cite{vauclair82}).
The resulting atmospheric abundances depend on the competition between
element diffusion and macroscopic motions like convection, mass loss,
rotation-induced mixing and other kinds of mixing processes.
Among the peculiar A stars, those which have large organized
magnetic fields (the so-called Ap stars) are more complex 
than the others (like Hg-Mn stars) : they show evidence of abundance spots or rings modulated by the magnetic structure. Among the coolest
Ap stars (effective temperatures between 6700K and 8700K) some 
oscillate with periods of a few minutes (the roAp stars) 
while the others seem stable (noAp stars). 
In roAp stars, the amplitudes of the oscillation modes are modulated
according to the rotation period, which is explained in the framework of
the
oblique pulsator model (Kurtz \cite{kurtz90}).

The helium $\kappa$-mechanism was invoked in the past as a possible
way to trigger the oscillations : a well-adjusted competition between helium settling and 
mass loss could increase the helium abundance in stellar
atmospheres as observed in helium-rich stars (Vauclair \cite{vauclair75}, 
Vauclair, Dolez {\&} Gough \cite{vauclair91}).
However, recent papers (Dziembowski {\&} Goode \cite{dziembowski96}, 
Balmforth et al. \cite{balmforth01}) show that, even in this case, helium is not able
to destabilize the star. On the other hand they find that,
in some cases, the oscillations in A stars can be
triggered by the hydrogen 
$\kappa$-mechanism. 
Such a process may be enhanced
by the settling of helium, which diffuses downwards in the absence
of strong mass loss.

In the present paper we study the asteroseismic signature of a helium
gradient 
inside main-sequence A-type stars. 
As mentioned by Gough (\cite{gough90}), rapid variations of the sound
velocity inside a star 
lead to partial reflections of the sound waves, which may clearly appear
as frequency 
modulations in the so-called ``second differences" : 
\begin{equation}
\delta_2\nu_{n, l} = \nu_{(n+1), l}+\nu_{(n-1), l}-2\nu_{n, l}
\end{equation}
which are computed 
for the same value of the
azimutal number $l$.
The modulation period of the oscillations is twice the ``acoustic
depth" of the region where the feature occurs (i.e. the time
needed for the sound waves to travel between this region and the
stellar surface).

Such an effect has been extensively studied in the literature for
stellar models in which helium settling is neglected. In this case,
the modulations in the frequencies are mostly due to the HeII
ionisation zone and to the edge between the convective and radiative
zones (Monteiro {\&} Thompson \cite {monteiro98}, Roxburgh {\&}
Vorontsov \cite {roxburgh01}, Mazumdar {\&} Antia \cite {mazumdar01},
Miglio et al. \cite {miglio03})
 
We show here that the presence of a diffusion-induced
helium gradient leads to a kink in the sound velocity with a very clear
signature in the oscillation frequencies : asteroseismic
observations of A-type stars can test 
helium diffusion and lead to a
precise value of the acoustic depth corresponding to the position of the 
helium gradient inside the star.

For the present study, we neglect the effects of magnetic fields
(the stars are assumed spherically symetric) and we do not compute the
diffusion of heavy elements. 
Complete computation of element diffusion including the radiative
acceleration on metals is a demanding work, especially for these stars
where heavy element diffusion occurs in the atmosphere, so that
radiative transfer should be solved in detail in the optically thin regions.
This is out of the scope of the present paper, where we are interested
in the structural changes induced by helium settling below the stellar 
surface and its influence on the oscillation frequencies.
It has recently been shown (e.g. Richard, Michaud {\&} Richer 
\cite{richard01}) that in case of pure diffusion iron should accumulate
deeper inside the stars, at temperatures around 200 000 K, where it can 
create a small convective zone.
As the results we obtain here are very encouraging, new computations including
heavy element diffusion should be done in the near future
to see whether such a feature may be checked with asteroseismology.

We apply this test
to the only roAp star in which enough modes have been observed, namely 
HD 60435 (Matthews et al. \cite{matthews87}).
The results are extremely encouraging
and show the importance of detecting as many modes 
as possible in rapidly oscillating Ap stars.

\section{Theoretical discussion}

\subsection{Seismic signatures of helium gradients}

Stellar acoustic $p$ modes with low $l$ values can propagate deeply 
inside the stars. For this reason, they may be used to obtain
information
on the deep stellar structure. 
However, in the case of strong gradients in the sound velocity, 
which may be due to the boundary of a convective zone, to 
the helium ionization region, or to helium gradients, the waves are 
partly reflected : this creates modulations in the frequency 
values which may be clearly visible in the 
second differences.
The modulation periods are 
equal to $ 2 t_s$ where $t_s$ is the time needed for
the acoustic waves to travel between the surface and the
considered region (acoustic depth), i.e. :
\begin{equation}
t_{s}=\int_{r_s}^{R}\frac{dr}{c(r)}
\end{equation}
where $c(r)$ is the sound velocity at radius $r$, and
$r_s$ the radius of the considered region. 

As the $l$ dependence of the oscillatory signal is small for high
frequency modes of low degrees (Monteiro, Christensen-Dalsgaard {\&}
 Thompson \cite{monteiro94}), computational results for modes of 
different low $l$ values can be treated together. 
 Here we show that the presence of
diffusion-induced helium gradients lead to very clear signatures
in the modulation of the second differences.

\subsection{Models with helium diffusion}

We have chosen to study the evolution of 1.6 M$_{\odot}$  
and 2.0 M$_{\odot}$  stellar models, 
with pure helium settling.
The stellar evolution code we use is the ``Toulouse-Geneva code'' 
which has been described many times in the literature (e.g. Richard et al.
\cite{richard96}, Th\'eado {\&} Vauclair \cite{theado03}).
The physical input and parameters are those described in Richard, Th\'eado
{\&} Vauclair (\cite{richard04}): they include the most recent studies
available for the equation of state, opacities and nuclear reaction rates.
Helium diffusion is treated as described in these papers (see also
Vauclair \cite{vauclair03}).
Metal diffusion is not included here
and no mixing process or mass loss 
is taken into account. 
In real stars, there must be some mixing which prevents element diffusion
from producing extreme abundance anomalies. Computations including
mixing as well as metal diffusion will be done in the near future.
The aim of the present paper is to study
the precise signature on the oscillation frequencies 
of helium gradients inside stars.

\begin{table}[t]
\catcode `\*=\active \def*{\phantom{0}}
\baselineskip12pt
\pagestyle{empty}
\centerline{TABLE 1}
\medskip
\centerline{Model parameters}
\bigskip

\halign{# & # & # & # & # & # & # & # \cr
\noalign{\hrule}
\noalign{\kern2pt}
\noalign{\hrule}
\noalign{\bigskip}
& mass (M$_{\odot}$) & age (Myrs) & *log($L/L_{\odot}$) & ***$T_{eff}$ & *****$t_{x}$(sec) & ****$t_{s}$(sec) & ****$r_{s}/R$ \cr
\noalign{\medskip}
\noalign{\hrule}
\noalign{\bigskip}
& 1.6* & **95 & ***0.84 & ***7936 & *******5400 & ******1480 & *****0.93 \cr
\noalign{\medskip}
& 1.6* & *591 & ***0.88 & ***7585 & *******6920 & ******2700 & *****0.86 \cr
\noalign{\medskip}
& 1.6* & 1600 & ***1.05 & ***7071 & ******11240 & ******5160 & *****0.80 \cr
\noalign{\medskip}
& 1.6h & 1600 & ***1.06 & ***7352 & ******10227 & ********** & ********* \cr
\noalign{\medskip}
& 2.0* & **63 & ***1.22 & ***9521 & *******5917 & ******1470 & *****0.94 \cr
\noalign{\medskip}
& 2.0* & *649 & ***1.31 & ***8293 & ******10150 & ******4100 & *****0.85 \cr
\noalign{\bigskip}
\noalign{\hrule}
}
\smallskip

Note.- $t_{x}$ represents the total acoustic radius of the models, $t_{s}$  
the acoustic depth at the location of the helium gradients 
and $r_{s}/R$ the corresponding fractional 
radii ; all the models presented in this table
include helium diffusion,
except the model labelled 1.6h which is homogeneous in its outer
layers.

\end{table}

We have computed the 
oscillation frequencies and their second differences in five models :
1.6 M$_{\odot}$ with ages 95 Myrs and 591 Myrs, and 1.6 Gyrs and 
2.0 M$_{\odot}$ with ages 63 Myrs and 649 Myrs (Table 1).
The results are given in Figs. 1 to 5. Only modes of degrees 
$l=0,1,2$ and $3$ have been taken into account
so that they could be treated together for the study of 
the second differences. The oscillation
 frequencies have been limited to the range 0.5 to 2 mHz to take into account
 the asymptotic approximation validity and the stellar cut-off. 
In this range 80 to 100 second differences are computed, according to the model.

In each figure six graphs
are displayed : a) the helium profile inside the star ; b) the
sound velocity, which shows a clear kink at the place of the
helium gradient ; c) the first derivative of the sound 
velocity in which the kink is still clearer ; d) the
second differences which show periodical oscillations ; 
e) the Fourier transform of these oscillations, in which
clear peaks are found for precise time values ; f) the
time needed for the acoustic waves to travel between the
surface and the considered radius, or ``acoustic depth".

We obtain clear signatures in the second differences
of the frequencies of the diffusion-induced helium 
gradients inside stars. It can be checked that
the periods corresponding to the peaks in the Fourier transforms (graphs e)
are exactly
twice the acoustic depths of the helium gradients (graphs f), as expected.
In these figures, the scales are all the same except for graphs d) and
f). 
In particular the Fourier transforms (graphs e) are presented at 
the same scale in each figure to 
show how the peak amplitudes decrease for deeper layers (larger times).
As shown by Mazumdar {\&} Antia (\cite {mazumdar01}), the amplitude of the
oscillatory signal in the second differences contains an amplification
factor of $4 $sin$^2(\pi t_s/t_*)$ where $t_*$ is the total acoustic
radius of the star and $t_s$ the acoustic depth of the partial reflection 
region. Here this factor increases from about 2 to about 4 as the models
evolve. The amplitude decrease by a factor $\simeq10$ obtained in
the present computations must be due to a competing process. 
It may be due partly
 to the fact that the energy in $p$ waves 
decreases towards the center and partly to the fact
that the helium gradient becomes smoother with age,
so that the reflected part 
(oscillations) is smaller for deeper discontinuities.

With these figures we can clearly follow the signatures 
of the helium gradients as they 
sink into the stars. The corresponding ages are obtained with the
assumption of pure diffusion. In real stars macroscopic effects
may change the time scales for helium settling. Comparisons with
observed frequencies will give the positions of helium
gradients inside the star, but we must keep in mind that 
the time needed for these
gradients to develop depends on the physics.

\begin{figure}
 \resizebox{\hsize}{!}{\includegraphics{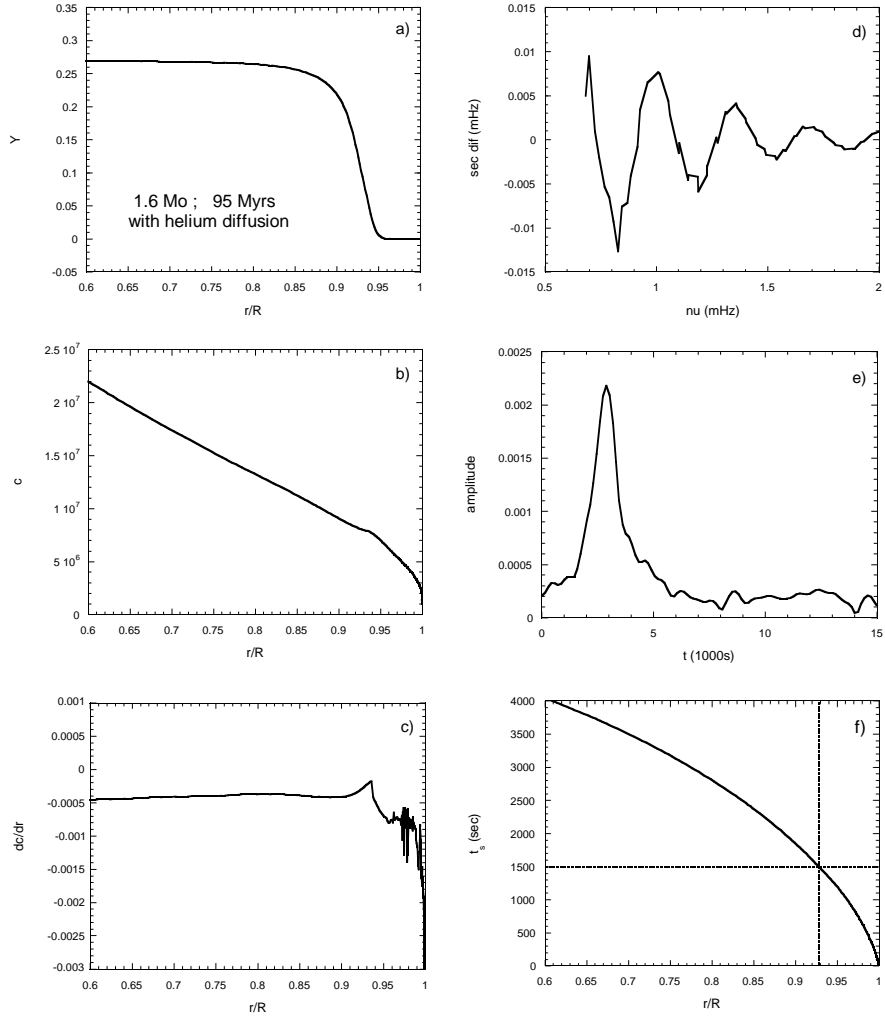}}
 \vspace*{1cm}
\caption{Helium diffusion and its consequences for the stellar 
structure and oscillation frequencies for a 1.6 M$_{\odot}$ 
at 95 Myrs ; a) helium profile as a function of the fractional radius ;
b) sound velocity : a kink is clearly visible at the place of the
helium gradient ; c) first derivative of the sound velocity : 
the kink is still more visible ; d) the second differences of
the oscillation frequencies plotted as a function of the
frequencies; e) the Fourier transform of graph (d) plotted as
a function of time (in thousands of seconds) ; 
the vertical dashed line corresponds to 
twice the total acoustic depth of the star (see Table 1) ;
f) the ``acoustic depth", or time needed for the acoustic waves
to travel from the surface to the considered radius : it can
be checked that
the peak of graph (e) corresponds to a time scale which 
is twice the ``acoustic depth" of the helium gradient.}
 \label{fig1}
\end{figure}

\begin{figure}
 \resizebox{\hsize}{!}{\includegraphics{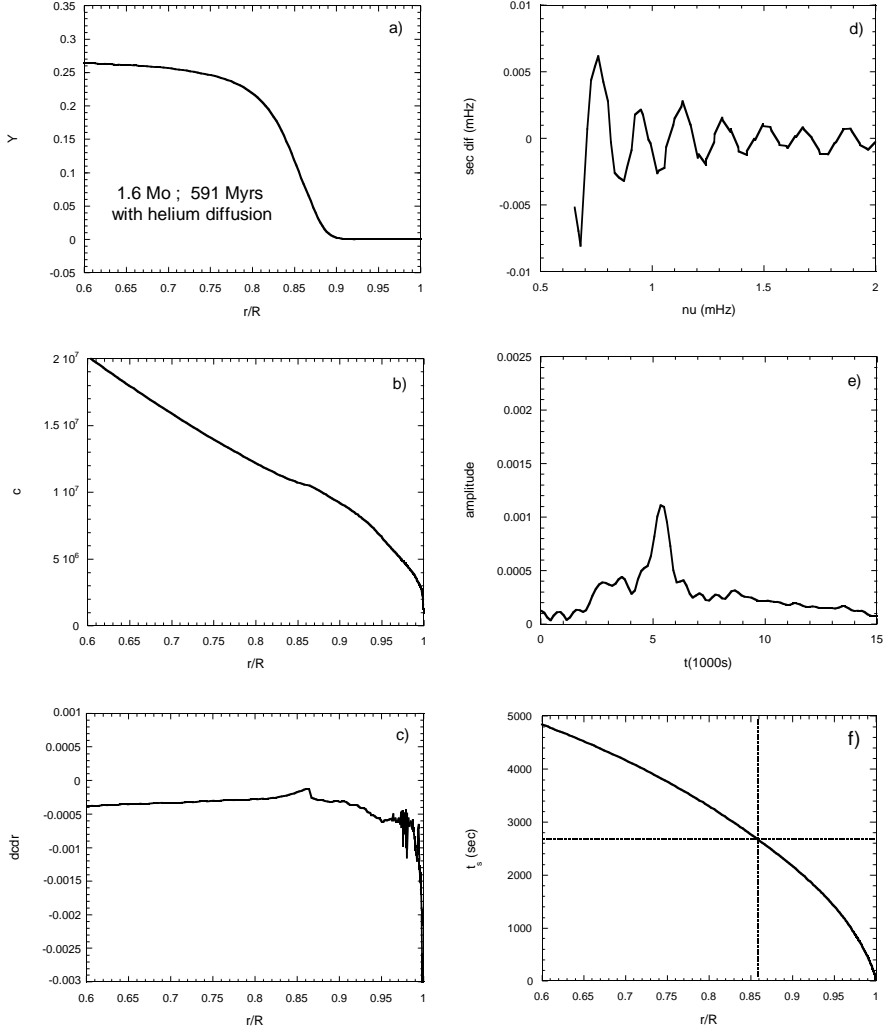}}
 \vspace*{1cm}
\caption{Same as Fig. 1, for a 1.6 M$_{\odot}$ model  at
591 Myrs.}
\label{fig2}
\end{figure}

\begin{figure}
 \resizebox{\hsize}{!}{\includegraphics{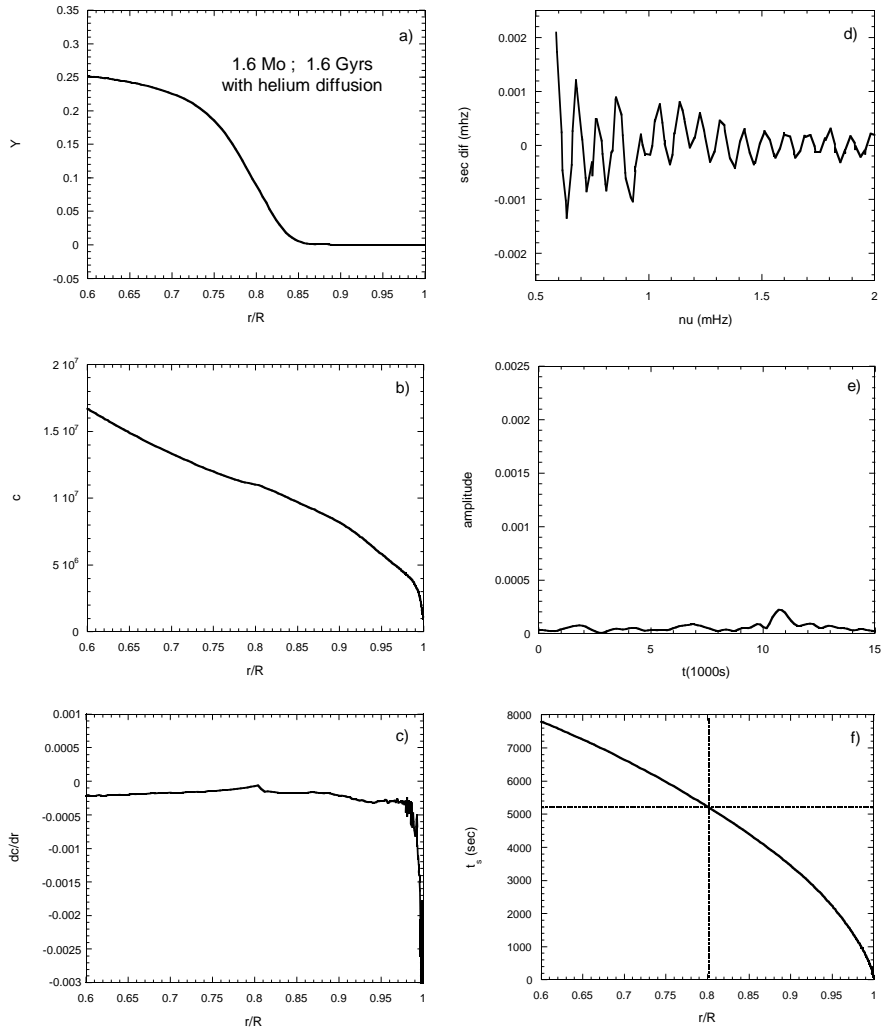}}
 \vspace*{1cm}
\caption{Same as Fig. 1, for a 1.6 M$_{\odot}$ model 
at 1.6 Gyrs.}
\label{fig3}
\end{figure}

\begin{figure}
 \resizebox{\hsize}{!}{\includegraphics{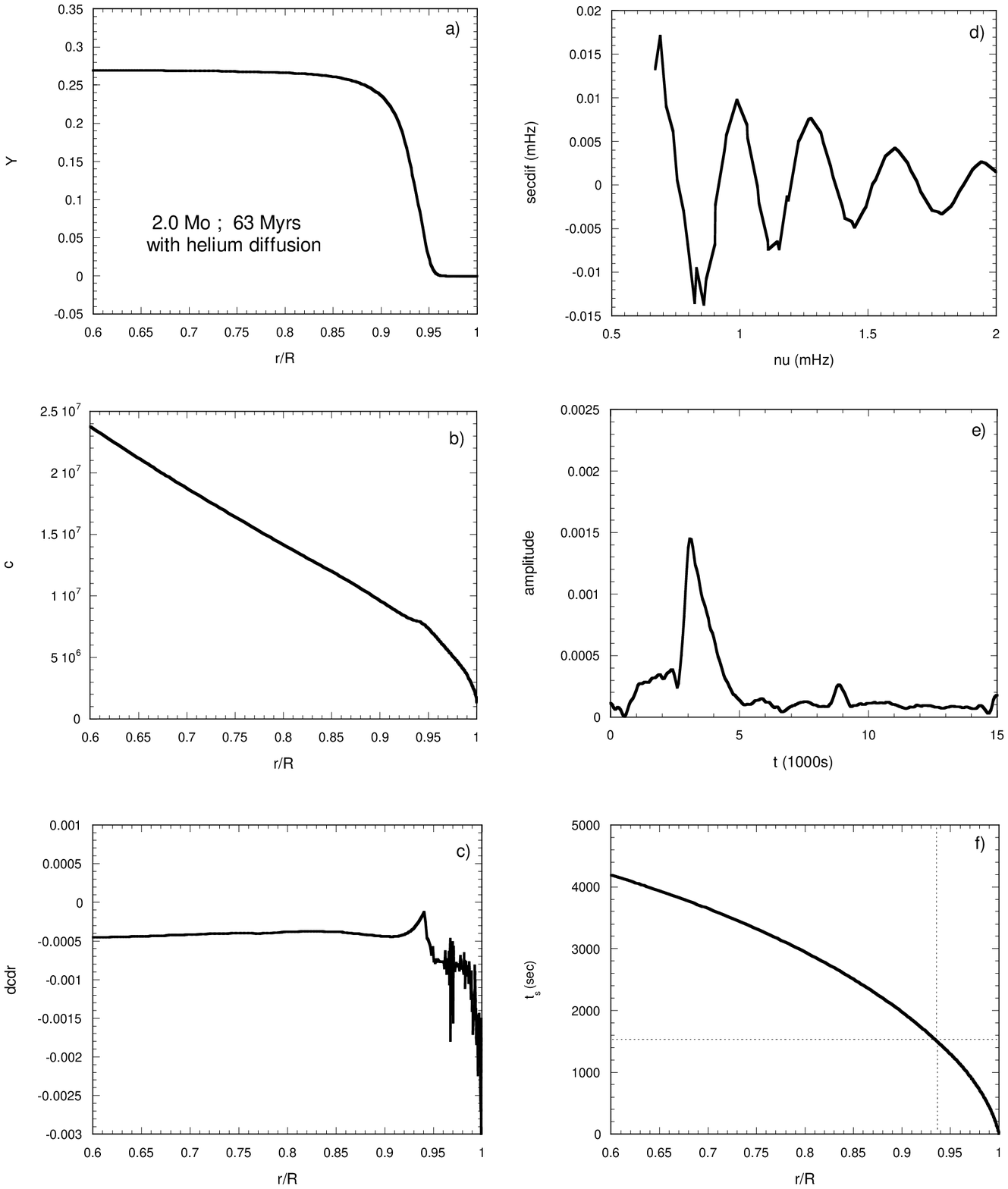}}
 \vspace*{1cm}
\caption{Same as Fig. 1, for a 2.0 M$_{\odot}$ model 
at 63 Myrs.}
\label{fig4}
\end{figure}

\begin{figure}
 \resizebox{\hsize}{!}{\includegraphics{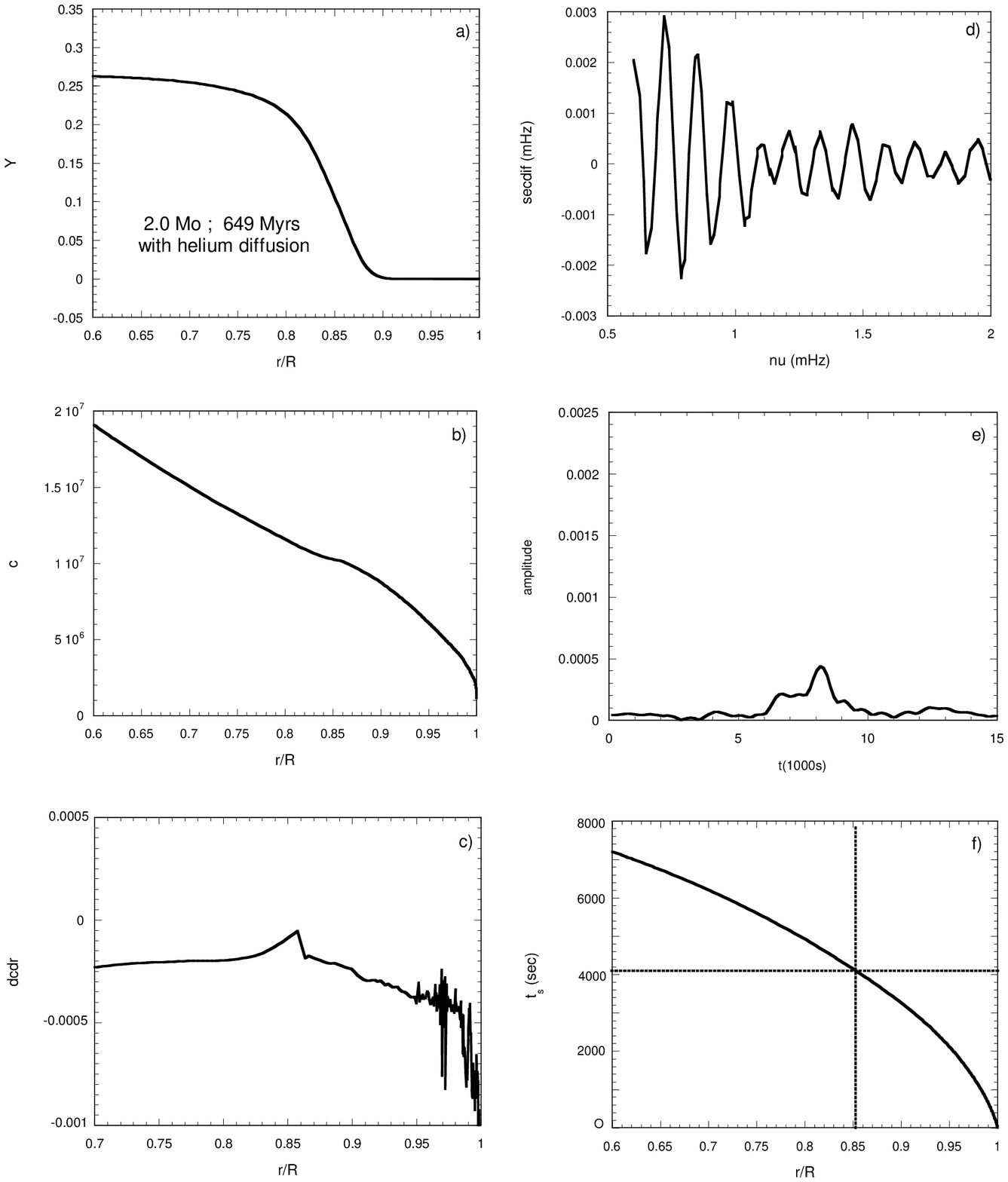}}
 \vspace*{1cm}
\caption{Same as Fig. 1, for a 2.0 M$_{\odot}$ model 
at 649 Myrs.}
\label{fig5}
\end{figure}

\subsection{Models without element diffusion}

Similar computations have been done for evolution models in which
helium diffusion is suppressed. 
In this case no helium gradient exists in the stellar outer layers 
and the oscillatory signal described in the previous section is
not present. However, helium ionization induces a typical
variation of the $\Gamma_1$ coefficient which gives rise to another
type of oscillatory signal in the frequencies.
Computational results are shown in Fig. 6 for the case of the 
1.6 M$_{\odot}$ model  at
1.6 Gyrs. In this case graph a) displays the 
$\Gamma_1$ coefficient in the stellar outer layers. 
As already studied by several authors (e.g. Miglio et al.
\cite {miglio03}), this
coefficient is strongly influenced by the helium ionisation
gradients.
We can see the corresponding kinks in the
sound velocity 
and its first derivative. 
This oscillatory signal does not appear when diffusion is effective as helium
is then depleted in the region considered.
Note 
that in the graphs of Fig. 6 the scales are quite different from those
of Figs. 1 to 5. The effect due to the helium ionization gradients is
much smaller than the effect due to the overall helium gradient in the case
of diffusion. 
Here the amplification factor  $4 $sin$^2(\pi t_s/t_*)$ is about 10 times
smaller for the homogeneous model than for the model of the same age
which includes
 helium diffusion. The large difference
 between the signal amplitudes in the two cases 
may be due to this effect.
The right-hand graphs display the second differences, 
their Fourier transform and the travel time of the waves, as in Figs. 1 to 5.
The features are very different from those obtained in
case of diffusion, where the signature of the helium gradient is
dominant.

\begin{figure}
 \resizebox{\hsize}{!}{\includegraphics{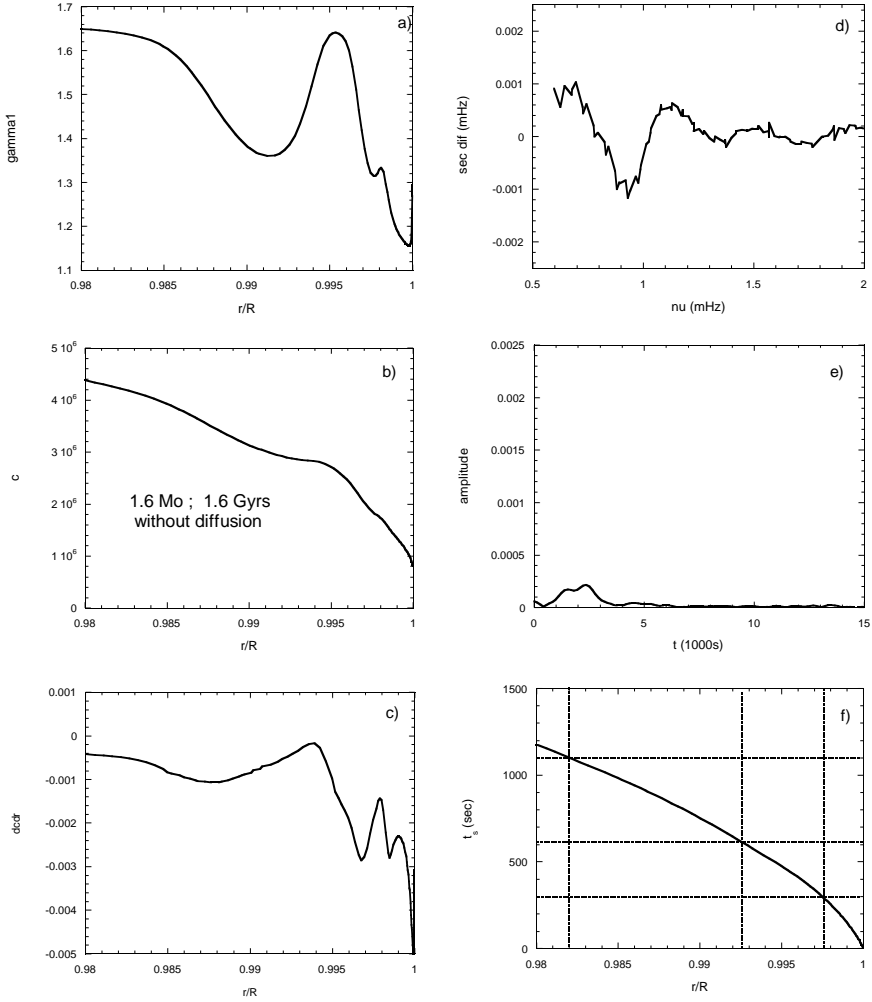}}
 \vspace*{1cm}
\caption{This figure is similar to Fig. 1 when element diffusion is
suppressed, 
in a 1.6 M$_{\odot}$ at 95 Myrs ; here graph a) represents the 
$\Gamma_1$ coefficient which shows a clear feature at the place
of the HeII ionisation region ;
the other
graphs
are similar to those of Fig. 1 ; we can clearly see the influence of the
helium ionisation regions
on the frequencies. Note that the effect is much smaller than
that of a helium gradient when diffusion is taken into account.}
 \label{fig6}
\end{figure}

\section{The roAp star HD60435}

Among all the rapidly oscillating Ap stars which have been detected
up to now, 
only one, HD60435, presents a frequency spectrum rich 
enough for the second differences to be computed.
This star has been extensively by Matthews et al (\cite{matthews87})
who discovered 17 frequency peaks in the fourier spectrum, which they
basically identified as $l = 1$ and $l = 2$ modes. As for all
roAp stars, the amplitudes of the oscillations are modulated
according to the rotation period (here 7.7 days). However,
contrary to most Ap stars, no magnetic field has yet been
clearly detected.

We have computed the observed second differences for this star,
using the frequencies given by these authors in their Table 2 (with
an uncertainty of $\pm 0.0001$mHz) and their tentative
identification, as given in the same table. For the $0.8428$ mHz frequency,
we chose the first identification given, i.e $(n-12, 2)$ and we assumed
that the $1.2848$ mHz frequency, which is not clearly identified in their
paper, corresponds to $(n-3, 1)$.

We present these frequencies and mode identifications
in Table 2, with separate columns
for $l = 1$ and $l = 2$. The second differences can only
be computed when at least three consecutive modes have been
identified for a given $l$. Table 3 gives the computed
second differences for HD60435 and the $l$ value of
the corresponding mode. The obtained
values for both $l$ are presented in Fig. 7. 
The modes are labelled with their $l$ value.

The second
differences clearly show a modulated trend similar to those
presented in Figs 1 and 4, characteristic of the presence of a 
helium gradient below the surface. The fourier transform of the observed
second differences is presented in Fig 8 (thick curve) together
with those computed for 1.6 M$_{\odot}$ (95 Myrs) and 2.0 M$_{\odot}$
(63 Myrs).
We can see that the observed
peak for HD60435 lies at a position close to those of the two models.
The maximum is found at a period of 3400 sec, which corresponds to an
acoustic depth of the helium gradient $t_s \simeq 1700$ sec.
With a spectral classification A3V, this star should have a mass
close to 2.0 M$_{\odot}$.We can deduce from this study that 
the modulation observed in the second differences for the star HD60435
may be a signature of a helium gradient at a radial depth of about $90\%$
to $95\%$ of the stellar radius.

\begin{table}[t]
\catcode `\*=\active \def*{\phantom{0}}
\baselineskip12pt
\pagestyle{empty}
\centerline{TABLE 2}
\medskip
\centerline{Observed frequencies and mode identifications in HD60435}
\bigskip

\halign{# & # & # & # & # & # \cr
\noalign{\hrule}
\noalign{\kern2pt}
\noalign{\hrule}
\noalign{\bigskip}
& $\nu$ (mHz) & mode id.& : & $\nu$ (mHz) & mode id. \cr
\noalign{\medskip}
\noalign{\hrule}
\noalign{\bigskip}
& 0.7090 & (n - 14, 1) & : &  0.8428 & (n - 12, 2)\cr
\noalign{\medskip}
& 0.7614 & (n - 13, 1) & : &  0.9397 & (n - 10, 2)\cr
\noalign{\medskip}
& ****** & ****** & : &  0.9906 & (n - 9, 2)\cr
\noalign{\medskip}
& 1.1734 & (n - 5, 1) & : &  1.0433 & (n - 8, 2)\cr
\noalign{\medskip}
& 1.2250 & (n - 4, 1) & : &  1.0990 & (n - 7, 2)\cr
\noalign{\medskip}
& 1.2848 & (n - 3, 1) & : &  1.1482 & (n - 6, 2)\cr
\noalign{\medskip}
& 1.3281 & (n - 2, 1) & : &  1.3525 & (n - 2, 2)\cr
\noalign{\medskip}
& 1.3810 & (n - 1, 1) & : &  1.4073 & (n - 1, 2)\cr
\noalign{\medskip}
& 1.4334 & (n , 1) & : &  1.4572 & (n , 2)\cr
\noalign{\bigskip}
\noalign{\hrule}
}
\smallskip

Note.- after Matthews et al 1987 ; 
left columns : $l=1$ 
right columns : $l=2$ .

\end{table}

\begin{table}[t]
\catcode `\*=\active \def*{\phantom{0}}
\baselineskip12pt
\pagestyle{empty}
\centerline{TABLE 3}
\medskip
\centerline{Computed second differences for HD60435}
\bigskip

\halign{# & # & # & # \cr
\noalign{\hrule}
\noalign{\kern2pt}
\noalign{\hrule}
\noalign{\bigskip}
& $\nu$ (mHz) & sec.dif. (mHz) & $l$ \cr
\noalign{\medskip}
\noalign{\hrule}
\noalign{\bigskip}
& 0.9906 & *+*0.0018 & 2 \cr
\noalign{\medskip}
& 1.0433 & *+*0.0030 & 2 \cr
\noalign{\medskip}
& 1.0990 & *--*0.0065 & 2 \cr
\noalign{\medskip}
& 1.2250 & *+*0.0082 & 1 \cr
\noalign{\medskip}
& 1.2848 & *--*0.0016 & 1 \cr
\noalign{\medskip}
& 1.3281 & *+*0.0096 & 1 \cr
\noalign{\medskip}
& 1.3810 & *--*0.0005 & 1 \cr
\noalign{\medskip}
& 1.4073 & *--*0.0049 & 2 \cr
\noalign{\bigskip}
\noalign{\hrule}
}
\smallskip

\end{table}

The uncertainties in the frequency
measurements lead to relative uncertainties of the
second differences smaller than $10\%$. 
Simulations of the error in the position of the amplitude peak
due to the uncertainties in the frequencies
show that the results are quite robust : the time $t_s$ at the
maximum does not change by more than $0.05\%$. A more important
uncertainty arises from the small number of observed frequencies.
A reasonable estimate of this uncertainty is obtained by assuming 
that a new maximum could appear 
between the presently observed one and the two adjacent points
if more frequencies were observed. Such an estimate leads to 
$\pm300$ sec in the Fourier transform, or : $t_s = 1700 \pm 150$
sec.

These results relie
on the mode identification which is given as ``tentative"
by Matthews et al (\cite{matthews87}). Note however that
the difficulty in the identification was due to the fact that the 
frequencies were not exactly at the expected values. 
The present study shows that the observed frequencies
are consistent with
the presence of a helium gradient.

\begin{figure}
 \resizebox{\hsize}{!}{\includegraphics{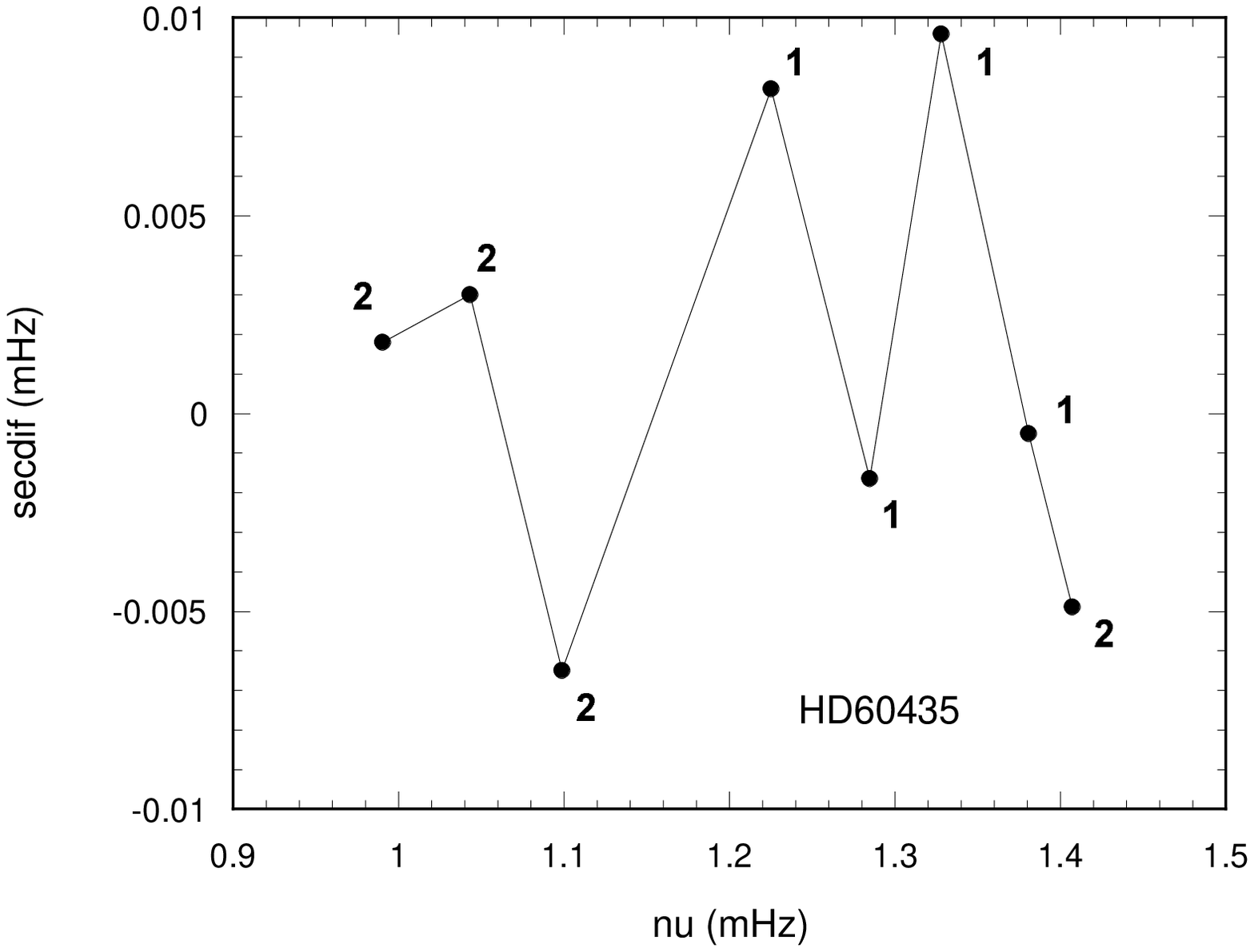}}
 \vspace*{1cm}
\caption{Second differences in the oscillation frequencies of HD60435
as found from Matthews et al. (1987) ; the points are labelled with their
$l$ values.}
 \label{fig7}
\end{figure}

\begin{figure}
 \resizebox{\hsize}{!}{\includegraphics{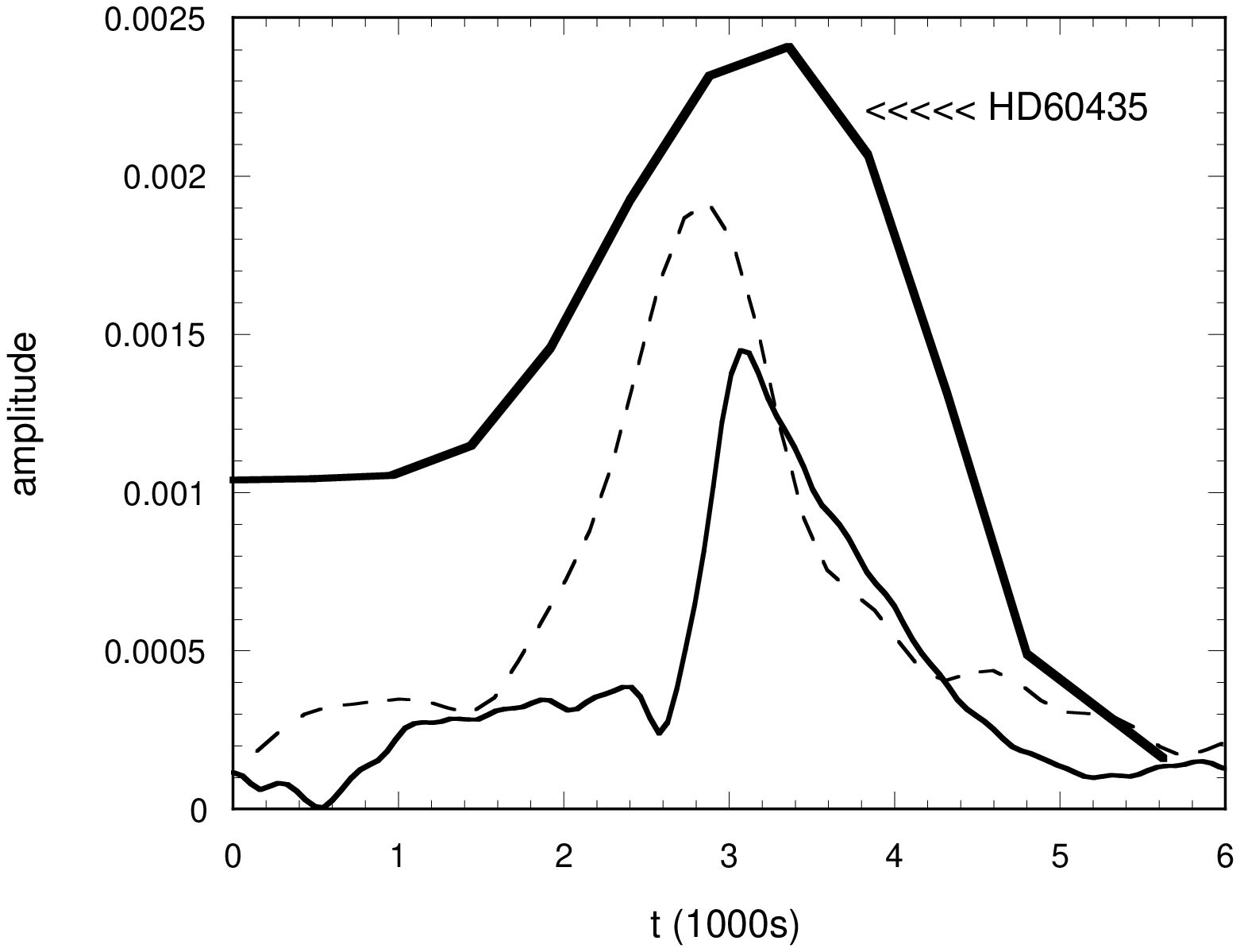}}
 \vspace*{1cm}
\caption{Fourier transform of the second differences of HD60435 (Fig.
7)
in heavy solid line ; the dashed curve represents the computed fourier
transforms of
the 1.6 M$_{\odot}$, $95$ Myr model and 
the light solid curve represents that of the
2.0 M$_{\odot}$, $63$ Myr model.} 
 \label{fig8}
\end{figure}

\section{Conclusion }

We have computed models of 1.6 M$_{\odot}$ and 2.0 M$_{\odot}$
main sequence stars with and without helium diffusion
and we have shown that asteroseismology can give clear
signatures of the presence of helium gradients 
below the surface in these A-type stars. The
analyses of the second differences in the oscillation frequencies
show modulations which are due to the partial
reflection of the acoustic waves at the place
of the helium gradient. 

At the present time these theoretical results can be
compared to the observations of only one star, 
namely the roAp HD60435. The computations
of the second differences can only been done
if enough modes with adjacent values of $n$
are detected for the same value of $l$. For
HD60435, with the presently detected modes,
we could compute 8 second
differences, which enabled us to give
evidence of a modulation similar to
the one expected in case of the presence of a helium
gradient below the surface. Furthermore, we could derive
the acoustic depth of this gradient : $1700 \pm 150$ sec.
The corresponding fractional radius is model-dependent, but
from our computations we suggest that it should lie 
around $r/R \simeq 0.9$ to $0.95$. 

In our evolutionary sequences, which are
computed with pure microscopic diffusion, a helium gradient at this
depth is obtained early in the stellar life (65 Gyrs).
However, in real stars microscopic diffusion must 
compete with macroscopic motions so that the helium
gradient can develop later in the evolution : our strong result
for HD60435 consists in determining the acoustic 
depth of the helium
gradient, not the age of the star.

As for all roAp stars, the amplitude of the oscillations observed
in this star is modulated according to the rotation period, of 7.7 days.
This modulation is probably related to the presence of a magnetic field,
although no clear detection have been obtained. 
Our theoretical computations have been done for spherically
symetrical stars, without rotation-induced mixing and 
ignoring any magnetic effect. 
In the future,
the influence
of the magnetic structure of the stars
will have to be studied in detail.
New models will also be computed, including
the abundance variations of heavy elements (with computations
of radiative accelerations), which have been neglected here.

The results we have obtained are consistent with the idea 
that helium diffusion does
occur in roAp stars.
It has been suggested in the past that the oscillations
observed in these stars could be triggered by helium 
$\kappa$-mechanism ; such a process could occur only
if helium accumulated due to the combined effect of
diffusion and a hypothetical stellar wind (Vauclair,
 Dolez, Gough \cite{vauclair91}). More recently
Balmforth et al (\cite{balmforth01}) showed that such
a helium accumulation was unable to destabilise the stars,
while the triggering could be due to the hydrogen $\kappa$-
mechanism, enhanced in the polar regions in case of 
pure helium diffusion. The present results are strongly 
in favor of this model.

In the future, 
new models will be computed, which will include
the abundance variations of heavy elements (with computations
of radiative accelerations). The
influence 
of the magnetic structure of the stars
will also have to be studied in details.

It will be interesting to try to
detect as many modes as possible in roAp stars to make
similar comparisons with other stars. 
Few modes are generally detected in these stars,
probably due to mode trapping.
New instruments and techniques may help in detecting
other modes. The frequencies should be determined
with a precision of order $0.1$ $\mu$Hz for 
the second differences to be computed with a
relative uncertainty of order $10\%$
and the method described here to
be applicable.
HD60435 should
also be studied in more detail with the most recent techniques
to obtain a higher precision of the results.

\begin{acknowledgements}

We thank the referee whose interesting critical remarks
and comments 
strongly helped improving this paper.
We warmly thank Hiromoto Shibahashi and
Don.W. Kurtz for stimulating discussions.
We also thank Stephane Charpinet who provided
his code for the computation of
the oscillation frequencies and Noel Dolez who provided his slow
Fourier transform code.
Sylvie Vauclair acknowledges a grant from Institut universitaire
de France and Sylvie Th\'eado the grant POCTI/1999/FIS/34549
approved by FCT and POCTI, with funds from the European
Community programme FEDER.

\end{acknowledgements}

\end{document}